\documentclass{aa}
\usepackage{natbib}
\usepackage{graphicx}
\usepackage{txfonts}
\usepackage{url}
\begin{document}

\title{Automated search for galactic star clusters in large multiband surveys:
I. Discovery of 15 new open clusters in the Galactic anticenter region}

\author{S.E. \,Koposov\inst{1,2,3}, E.V. Glushkova\inst{2}\and
I.Yu. Zolotukhin\inst{2}}

\offprints{E. Glushkova, \email{elena@sai.msu.ru}}

\institute{Max Planck Institute for Astronomy, K\"onigstuhl 17, D-69117, Heidelberg, Germany
\and
Sternberg Astronomical Institute, Universitetskiy pr. 13, 119992 Moscow, Russia 
\and
Institute of Astronomy, University of Cambridge, Madingley Road, Cambridge
CB3 0HA, UK \\
S. Koposov \email{koposov@mpia.de}, E. Glushkova \email{elena@sai.msu.ru}
 }

\date{Received <date> /Accepted <date>}
\abstract{}{According to some estimations, there are as many as 100000 open clusters
in the Galaxy, but less than 2000 of them have been discovered, measured,
and cataloged. We plan to undertake data mining of multiwavelength
surveys to find new star clusters.}
{We have developed a new method to search automatically for star clusters in
very large stellar catalogs, which is based on convolution with density
functions. We have applied this method to a subset of the Two Micron All Sky
Survey catalog toward the Galactic anticenter.

We also developed a method to verify whether detected stellar groups are real
star clusters, which tests whether the stars that form the spatial density
peak also fall onto a single isochrone in the color-magnitude diagram. 
By fitting an isochrone to the data, we estimate at the same time the main
physical parameters of a cluster: age, distance, color excess.
}
{For the present paper, we carried out a detailed analysis of 88 overdensity
peaks detected in a field of $16\times16$ degrees near the Galactic
anticenter. From this analysis, 15 overdensities were confirmed to
be new open clusters and the physical and structural parameters were determined
for 12 of them; 10 of them were previously suspected to be open clusters by
Kronberger (2006) and Froebrich (2007). The properties were also
determined for 13 yet-unstudied known open clusters, thus
almost tripling the sample of open clusters with studied parameters in the
anticenter. The parameters determined with this
method showed a good agreement with published data for a set of well-known
clusters.}{}
\titlerunning{Automated search for star clusters in large surveys}
\authorrunning{Koposov et al.}

\maketitle

\section{Introduction}

Star clusters are unique laboratories for investigation of a wide range of
astrophysical problems relating to star formation, stellar evolution, the
formation and structure of the Milky Way, and the distance scale of
the Universe.  As star clusters are usually single-age and single-metallicity populations, distance, age, and reddening in the cluster's direction can be determined 
with much higher accuracy than for isolated, or "field", stars.
To define at least
a reliable ranking of the open cluster properties, we need a large sample of
objects whose age, distance, and metallicity are accurately and homogeneously
known. So far, 1756 open clusters were cataloged \citep{dias2002}, 
but the basic physical parameters are known for less than 700
objects. And, all these parameters were derived by different authors
based on heterogeneous observational data.
Most of the open clusters in the Galaxy are probably not yet found
because open clusters are concentrated near the Galactic plane where
extinction by interstellar dust is most severe. Some literature estimates
put the total number of open clusters in the Galaxy at $10^{5}$ 
\citep[see, for example][]{surdin2000}. Modern all-sky
surveys (e.g., Two Micron All Sky Survey (2MASS), Deep Near Infrared Survey
of the Southern Sky (DENIS), Sloan Digital Sky Survey(SDSS), etc.)
provide a large store of information to study open clusters comprehensively
and homogeneously. Near-infrared surveys are especially useful because the data 
are far less affected by high reddening in the Galactic plane where the most open 
clusters are located.

Numerous attempts were made in recent years to search for star clusters
using such large surveys. However, the total number of newly-discovered
clusters with robust determinations of their physical parameters does not exceed
two dozen. \citet{dutra2003} performed a visual search for
IR clusters and similar objects in the direction of known nebulae using the
2MASS Atlas and found 179 embedded clusters and stellar groups. However,
it proved impossible to find the physical
parameters of this type of objects through isochrone fitting. \citet{ivanov2002} and
\citet{borissova2003} found 11 peaks by automated algorithm
and 3 peaks by visual inspection in the apparent stellar surface density in
the 2MASS point source catalog. They detected mostly embedded IR clusters, so
the
physical parameters could be derived only for one object. \citet{drake2005}
performed an automated search for clusters in the United States Naval
Observatory (USNO) A2 catalog using the method
developed by \citet{ivanov2002} and found 8 new candidates. However, their
basic parameters were not derived. \citet{kronberger2006}
visually inspected Digitized Sky Survey (DSS) and 2MASS images and selected 66
candidate clusters.
For 9 of 24 of the most probable clusters within this sample, the authors
determined
fundamental parameters by isochrone fitting.
\citet{froebrich2007} used star density maps obtained from 2MASS and found 1021 new 
cluster candidates. The authors statistically evaluated the contamination of their sample 
to be of about 50\% and left verification of the nature of each individual
cluster for future investigations.

We developed a new efficient method of searching stellar catalogs for
star clusters of different radii based on the convolution of the cataloged
stellar
source density maps with Gaussians 
\citep[a similar method was used to search for
dwarf Spheroidal galaxies and globular clusters in SDSS;][]{gc_kop,lf_paper}.
The method automatically finds cluster candidates and then confirms them by
testing whether the spatially clustered, potential cluster members lie
on the same isochrone in the color-magnitude diagram. At the same time, this
procedure determines the basic cluster parameters (age, radius, distance, and
color
excess) by fitting the isochrone position.
Below, we describe our method of automatic search for stellar overdensities
and the results of its application to the 2MASS data in the field of $16\times16$
degrees in the region of the Galactic anticenter.

\section{Method of automated search for star clusters}
\label{sec:method}
This work aims to provide a fast, simple, and efficient method of 
identifying open clusters in very large photometric catalogs, such as
2MASS, SDSS, DENIS etc. The overall density of the stars in the
Milky Way (MW) is high at low latitudes and can vary rapidly because of 
dust, etc. So, it is not easy to find star clusters algorithmically on
such a complex background. Even when a peak is found, it is important to
check whether the candidate is indeed an evolutionally connected group of 
cluster members or merely a group of stars clustered by chance.

\begin{figure}
\resizebox{\hsize}{!}{\includegraphics{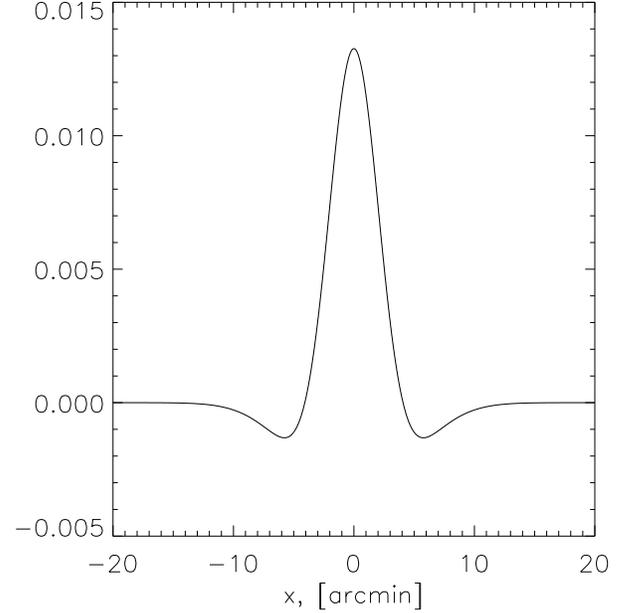}}
\caption{The one-dimensional slice of the 2D filter which we used for the
convolution for $\sigma_1=3\arcmin$ and $\sigma_2=6\arcmin$} 
\label{filter}
\end{figure}

Therefore, the method must be capable of detecting density peaks on a sharply
changing background and to evaluate their statistical significance. To develop a
universal technique for all data sources, we have to make it independent of any
pixelization effects and thus applicable to star structures of any size.
To ensure this, we built a density map, which presents the number of counts in
(RA, Dec) coordinates. Then, we convolved this image with a special filter
demonstrated in Fig.~\ref{filter}. The filter curve
is the difference between two 2-D Gaussian profiles and has zero integral.
Employing this special shape of filter, we ensure that a flat, or even
slowly changing background produces a zero signal, whereas the concentrations of
stars exhibit a high signal. The family of such filters called Difference of
Gaussians are well studied and used in computer vision science for feature
detection at various scales \citep{babaud,lindenberg}. The convolution
with this filter is equivalent to the subtraction of the density maps convolved
with the gaussians of different widths. The density map convolved with the small
Gaussian is used to detect the small scale overdensities, while the density map
convolved with wider Gaussian
is the estimation of local background. 

\begin{figure}
\resizebox{\hsize}{!}{\includegraphics{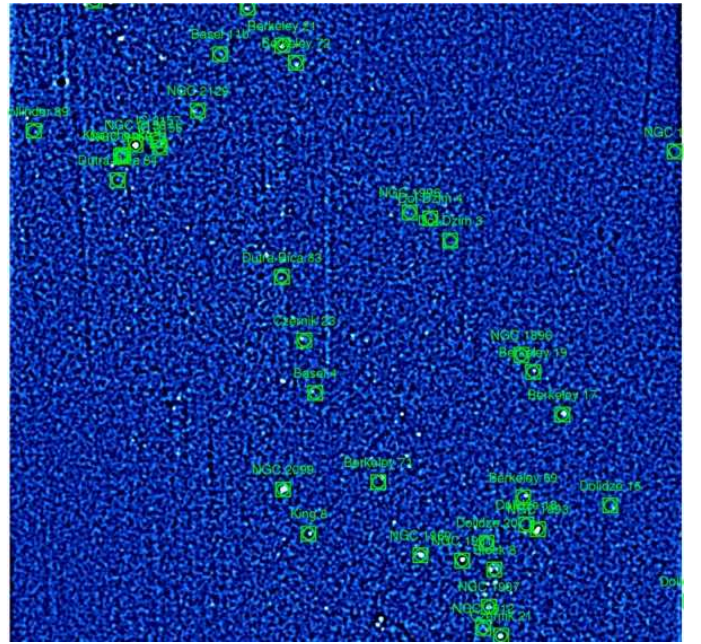}}
\caption{The $16\degr \times 16\degr$ map of the overdensities in the
anticenter region from 2MASS PSC.
The known open clusters from the Dias catalog \citep{dias2002} are marked.}
\label{anticenter_map}
\end{figure}

The following formulae demonstrate our convolution procedure.
First, we obtain the distribution of stars on the sky M(ra, dec):
$$M(ra, dec) = \sum\limits_i \delta(ra, dec)$$
Then, this map is convolved with the filter:
$$M(ra,dec) = M(ra,dec) * \left(G(ra,dec,\sigma_1) - G(ra,dec,
\sigma_2)\right)$$
where $G(ra,dec,\sigma)$ is the circular 2D Gaussian with unity integral and
width of $\sigma$.
On the last step, we normalize the convolved map to get the statistical
significances of all density fluctuations:
$$S(ra,dec) = \sqrt{4 \pi} \sigma_1 \frac{ M(ra,dec) * \left(G(ra,dec,\sigma_1)
-G(ra,dec,\sigma_2)\right)}{\sqrt{M(ra, dec) * G(ra,dec,\sigma_2))}}$$
That map S(ra,dec) shows the deviation of $M(ra,dec) * G(ra,dec,\sigma_1)$
above the background estimate given by $M(ra,dec) * G(ra,dec,\sigma_2)$.
Under the assumption of Poisson distribution of sources and $\sigma_2 \gg
\sigma_1$, the S(ra,dec) should be normally distributed with variance of 1.
The detection of overdensities on this map is very easy and may be done by
simple thresholding, i.e., to select all the overdensities more statistically
significant than $5\sigma$, we need find all the pixels on the map having
$S(ra,dec)>5$. We typically used the detection threshold of 4.5
sigmas.

Note that for clusters of the size close to that of the inner Gaussian, the
filter is very close to optimal. Also, it is important to understand that the
choice of $\sigma_2$ is related to the scale on which the background
estimate is obtained. Ideally, $\sigma_2$ should be rather large and much larger
than the $\sigma_1$, but unfortunately the 2MASS data in the MW plane
suffers significantly from extinction and the density of stars varies
on very small scales. Therefore, we are forced to use the $\sigma_2$, which is
not 
much larger than $\sigma_1$, to have a more local estimate of the background.
See also \citet{lf_paper} for a discussion of the  method.

The example of the convolved image from the 2MASS point source catalog for the
 $16\degr \times 16\degr$ anticenter region of our Galaxy is shown in
Fig.~\ref{anticenter_map}. A large
population of peaks is clearly seen. As we show below, most of these peaks can
be attributed to open clusters.

After the density peaks are detected, each individual peak should be
examined as to
whether it relates to a real cluster or just a random fluctuation. To
answer this question, we built the Hess-diagram representing the spatial density
of stars on a color-magnitude diagram (CMD) (The Hess-diagrams used
here are actually the difference of the 2D histogram of color-magnitude of
stars inside the circle around the overdensity center
(cluster) and the 2D histogram of color-magnitude of stars in the annulus
outside the circle (field)). As rule, a real
cluster is seen on this Hess-diagram by its main sequence and sometimes, by a
red giant branch. If we detected cluster on the Hess-diagram, we fitted its
CMD with the isochrone of solar metallicity taken from ~\citet{girardi2002}. To
do this, we automatically shifted the isochrone along the coordinate axes with
steps
equal to $1/100$ of the full range of ordinates and abscissas. At every
step, we varied the isochrone age in the interval of $log(age)$ from $6.60$ to
$10.25$ with the step of $0.05$. At each step, for each isochrone age, we
automatically built the radial density distribution for stars lying in the
vicinity of the isochrone (the distance in the color index is less than $0.05$ -
assumed cluster members) and for all other stars (the distance is greater than
$0.05$ - supposed field stars). Ideally, the distribution function for field
stars should be flat whereas the cluster members should feature a noticeable
concentration toward the center. In practice, the radial density distribution
of field stars shows a weak central concentration because of the number of
unresolved binary stars -- cluster members fall into this category (see, for
example, Fig.~\ref{radial_distrib_koposov52}) -- or because of the poor
photometry. That is why to qualify the radial density distribution, we
calculated the contrast ratio in the following manner: we divided the value of
the peak to the mean plateau for each distribution, then we found the ratio of
the two values obtained. For example, in Fig.~\ref{radial_distrib_koposov52} the
contrast is $(4.9/1)/(2.7/2)=3.63$. We believe that the best position of
the
isochrone on the color-magnitude diagram corresponds to the maximum contrast
ratio at radial density distribution. Testing our technique for well-studied
open clusters, we found
that the contrast ratio should be greater than 2. 
By fitting the position of the isochrone to obtain the maximum contrast ratio on
the density plot, we simultaneously found main physical parameters of a cluster:
age, distance, and color excess. If all plots (Hess-diagram, CMD, and the radial
density distribution) verified the reality of the cluster, then the overdensity
under study was considered to be a real cluster. Thus, we developed a method of
an automated search for stellar overdensities and proposed a reliable criterion
for verification of these overdensities as star clusters.

\section{Application of the method to the 2MASS data}

The 2MASS survey \citep{2mass} gives us 
a comprehensive dataset both to search for new open clusters and to
test our method: it covers 99.998\% of the sky with uniform precise photometry
and astrometry in the $J$ (1.25 $\mu$); $H$ (1.65 $\mu$); and $K_s$ 
(2.16 $\mu$) photometric bands. The global 2MASS sensitivity is 15.8 for
$J$-band; 15.1 for $H$-band; and 14.3 for $K_s$ at $S/N = 10$. For this reason,
we investigated the
$(J,J-H)$ color-magnitude diagrams in our work, but also used $(K_s,J-K_s)$
diagrams to confirm found cluster parameters and be sure that the relation
$E(J-H)/E(J-K_s)$ agrees with the normal extinction law.  

To extract $J$, $H$, $K_s$ photometry and astrometry data, we used the Virtual
Observatory resource named Sternberg Astronomical Institute Catalogue Access
Services (SAI CAS; \url{http://vo.astronet.ru}), allowing us to access the
largest astronomical
catalogs \citep{sai_cas}. 
For our purposes, we only selected the stars that have the quality flags
better than
U in each filter $J,H,K_s$.

Our primary goal was to detect clusters that have diameters from few to ten
arcminutes, so we used $\sigma_1=3\arcmin$ and $\sigma_2=6\arcmin$ in the filter
function. Other than the richest ones, clusters having diameters more than 15-20
arcminutes do not show as a rule the visual overdensity on the sky.
Such extended clusters are usually found by common proper motions or
radial velocities.

As a first application, we studied a field of 16 by 16 degrees towards the Galactic
anticenter and detected 88 density peaks of $>4.5\times \sigma$ significance. 
We compared these cluster candidates
with open clusters listed by \citet{dias2002} (in practice, we used
the online version of Dias' catalog at
\url{http://www.astro.iag.usp.br/~wilton/}) ; 23
of our significant peaks can be matched to known, optically-visible clusters. 
Furthermore, we matched 9 density peaks to embedded infrared clusters from the
list created by \citet{bica2003a,bica2003b}. Dias' catalog contains an
additional 15 open clusters in
this region, but our method does not detect them. Six of 15 clusters are not
reliable clusters: they are not found in the DSS and 2MASS images, three of them
are doubtful clusters according to \citet{dias2002}, one such object has no
entry in the WEBDA database on open clusters developed by E. Paunzen
and J.-C.Mermilliod (\url{http://www.univie.ac.at/webda/}).
Another 5 clusters having
diameters ranging from 20 to 60 arcmin according to \citet{dias2002}, are
considered to be clusters due to
the common pattern of the proper motion of star members, and only 2 clusters
among them, NGC 1912 and NGC 2168, exhibit stellar overdensities. The
remaining four clusters have density peaks under  $4.5\times \sigma$
significance, and three of them are seen on 2MASS images as weak embedded
clusters. These have no available data, except diameters, in Dias' catalog. 
Because more than half of the known clusters, that are on our list of detected
density peaks have unreliable or no parameter measurements, we completed the
detailed
analysis of all 88 peaks, including known clusters.

\begin{figure}
\resizebox{\hsize}{!}{\includegraphics{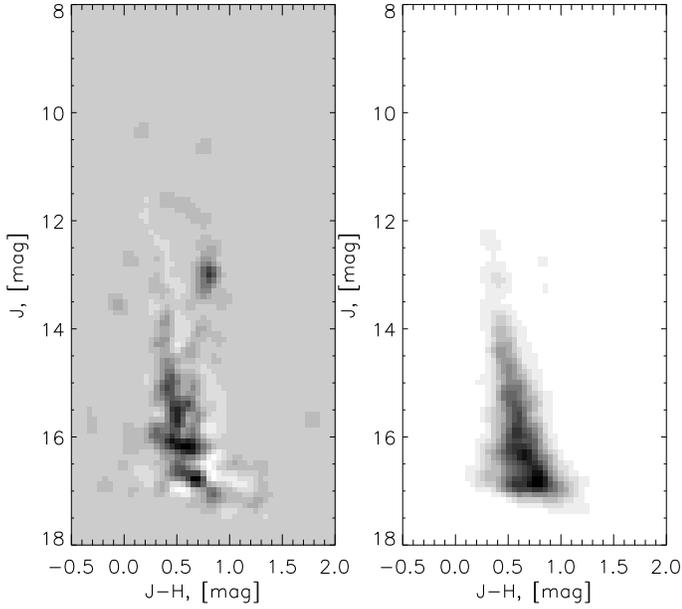}}
\caption{Hess-diagrams for open cluster Koposov 52. The left panel shows the
Hess diagram of the central 4\arcmin of the cluster with subtracted Hess diagram of the background.The right panel shows the Hess diagram of the background stars.}
\label{hess_koposov52}
\end{figure}

We built the Hess-diagram in $(J,J-H)$ coordinates. We plotted CMD within radius
$r$ with the value between 2 and 7 arcmin (depending on the cluster size) around
the overdensity center, then we subtracted the CMD for field stars built in
the ring between the two radii: $3\times r$ and $4\times r$. 
Each CMD was previously normalized
to the number of stars and smoothed using a 3-pixel Gaussian.
Figure~\ref{hess_koposov52} displays
the Hess-diagram for the new open cluster Koposov 52 built within the radius
of 4 arcmin around the cluster center (left-hand side) and CMD for field
stars in the ring around the cluster (right-hand side). The cluster can
clearly be seen on the Hess-diagram by its main sequence and red
clump stars. 

\begin{figure}
\resizebox{\hsize}{!}{\includegraphics{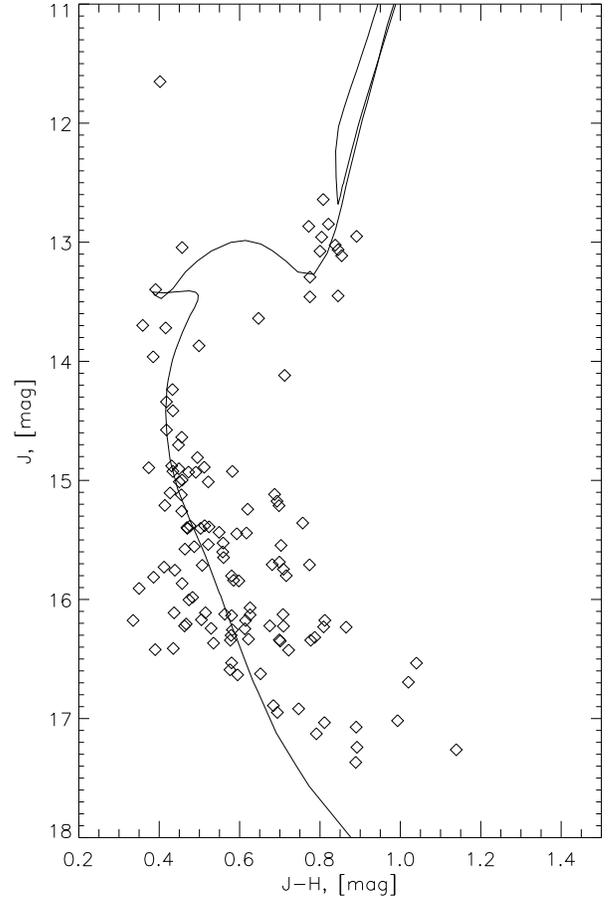}}
\caption{Color-magnitude diagram for cluster Koposov 52 inside the radius of 2 arcminutes. The fitted isochrone with $log(t)=8.95$, $E(B-V)=0.04$,  $(m-M)_0=12.32$ is overplotted.}
\label{cmd_koposov52}
\end{figure}

\begin{figure}
\resizebox{\hsize}{!}{\includegraphics{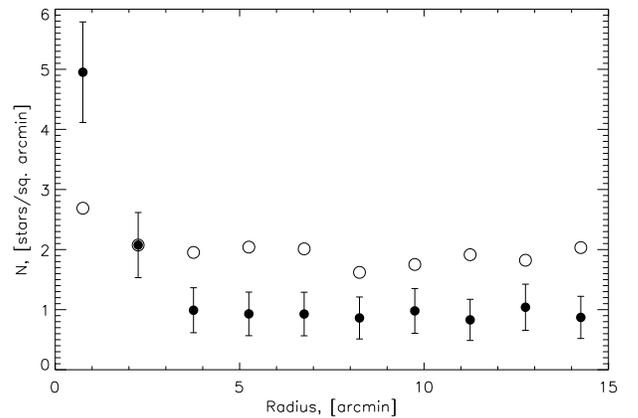}}
\caption{Radial density distribution for stars in the field of Koposov 52.
Solid circles represent the density of the stars lying closely to 
the isochrone. Open circles denote the density of the field stars 
lying far from the isochrone.}
\label{radial_distrib_koposov52}
\end{figure}

If a cluster-suspect manifested itself at the Hess-diagram, then we fitted its
CMD with an isochrone, by~\citet{girardi2002} of solar metallicity and
simultaneously verified whether the detected stellar group was a real star
cluster by plotting radial density distribution for stars lying on the isochrone
and for all other stars (see for details Sect.~\ref{sec:method}).
The fitted isochrone with the age of $log(t)=8.95$ for Koposov 52 is shown in
Fig.~\ref{cmd_koposov52}.
Star members are taken within the radius of 2 arcmin around its center; the
position of the isochrone leads to the following estimations: $E(J-H)=0.34$
and $(m-M)_J=13.20$. The radial density
distribution corresponding to this fitted isochrone is displayed in Fig.~\ref{radial_distrib_koposov52}:
solid circles denote the stars deviating from the isochrone by less than
0.05 magnitude in color $(J-H)$; open circles denote all other stars.
Therefore, the total density profile of all stars is the sum of the open
circle profile and solid circle profile. The errors plotted on the datapoints
are simple Poisson errors. The ratio of contrasts for the "isochrone" and
"field" stars calculated in Sect.~\ref{sec:method} equals to $3.63$. 
It is important to note that the background value (value at large radii) for the
solid circle profile is much lower than the background value for the empty
circle profile. This illustrates the advantage of using the isochrone filter for
the detection and analysis of overdensities -- with such a filter, 
overdensity is much more obvious due to the reduced background.

We performed isochrone fitting on $(J,J-H)$ diagram because there is a higher
magnitude limit for $J$-band in 2MASS, and used a 15-arcmin region around stellar
overdensity. We independently performed the same fitting procedure on a 
$(K_s,J-K_s)$
CMD and compared the distances obtained from two fittings with each other, and
the relation $E(J-H)/E(J-K_s)$ with the normal extinction law given by \citet{cardelli1989}, which equals to 0.55. Also we used the relations
$A_{K_s}=0.670\times E(J-K_s)$, $A_J=0.276\times A_V$, 
and $E(J-H)=0.33\times E(B-V)$ from the  paper by \citet{dutra2002}.

The cluster was considered to be a real cluster, if all plots (Hess-diagram,
CMD, and the radial density distribution) verified the reality of the cluster.

\section{Results}

We used the routine described above to study all 88 overdensities.
We found that 11 stellar overdensities turn out to be new, optically-visible
clusters. The parameters for all
these clusters are listed in Table~\ref{param_new_OC}.
One of them, Koposov 52,
was published earlier as KSE18 \citep{koposov2005,zolotukhin2006}
and then independently found by \citet{kronberger2006} as Teutsch 51.  
Four other clusters coincide with cluster candidates from the list compiled by 
\citet{froebrich2007}, and two clusters coincide with stellar
agglomerations reported by
\citet{kronberger2006}. Note that in both papers cited, these objects are
considered as possible clusters with futher investigation necessary to
clarify their nature. In our work, we not only discovered these clusters by an
independent method, but also verified their nature and found their fundamental
parameters. That is why we consider them as new clusters.

Note that we are not able to make a precise estimation of the age of the
young
clusters, if their color-magnitude diagrams do not exhibit red or blue
giants or supergiants. Provided that the positions of an isochrone with
respect to the axes of coordinates is unchanged, we can alter its age within
a broad range (in some cases, up to $8.20$ in $log(age)$) without having a
noticeable effect on the contrast on the radial-density distribution plot.
This can be attributed to the existence of an extended vertical section of
$(J,J-H)$ and $(K_s,J-K_s)$ isochrones and the absence of massive stars in poor
clusters. Therefore, we were only able to make the upper
boundary for the age estimation of Koposov 36 and Koposov 53, Koposov
10, Koposov 27, and Koposov 49 clusters. While it is possible that some of the
bright stars on the CMD are giants belonging the cluster, there is no
statistically meaningful way to check that with only photometric data, therefore
we assign only the upper limits for the ages of these 5 clusters.

The errors in color excess, distance
moduli, distances and ages are evaluated from the differences in the parameters 
derived from isochrones fitted in $(J,J-H)$ and $(K_s,J-K_s)$ diagrams. Hess-diagrams, 
radial density distribution, and fitted isochrone in $(J,J-H)$ CMD's for 
10 new clusters (except for Koposov 52) are given in
Fig.\ref{k10}-\ref{k77} in the Appendix. 
Four of the 11 new clusters are relatively young,
less than 100 Myr, whereas the other clusters in this set are very old,
exceeding 1 Gyr. Distances of all clusters from the Sun are ranged
between 1.5 and 3.5 kpc.

\begin{figure}
\resizebox{\hsize}{!}{\includegraphics{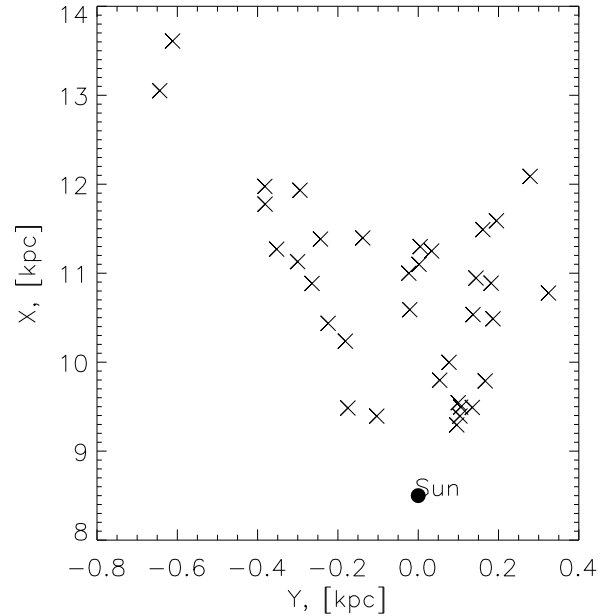}}
\caption{Distribution of clusters across the galactic plane. 
}
\label{clusters_distribution}
\end{figure}

\begin{table*}
\caption{Parameters of new clusters} 
\label{param_new_OC}
\centering
\begin{tabular}{llcccccccc}
\hline\hline
{Name} & 
{Other name} &
{RA (J2000)} & 
{Dec (J2000)} & 
{D} & 
{E(B-V)} & 
{$\frac{E(J-H)}{E(J-K)}$} & 
{$(m-M)_0$} & 
{Distance} & 
{Age} 
\\
 &
 &
{h:m:s} & 
{d:m:s} & 
{arcmin} & 
{mag} & 
 & 
{mag} & 
{pc} & 
{log(yr)}

\\
\hline
Koposov 10 & FSR 795 & 05:47:28.6 & +35:25:56 & 4 & 0.81$\pm$0.25 & 0.79 &
11.54$\pm$0.3 & 2000$\pm$300  & $<$8.6 \\
Koposov 12 & FSR 802 & 06:00:56.2 & +35:16:36 & 9 & 0.30$\pm$0.03 & 0.51 &
11.55$\pm$0.03 & 2050$\pm$50  & 8.90$\pm$0.1 \\
Koposov 27 & Teutsch 1 & 05:39:30.0 & +33:21:00 & 3 & 0.45$\pm$0.1 & 0.69 &
12.8$\pm$0.5 & 3700$\pm$900 & $<$8.65 \\
Koposov 36 & & 05:36:50.6 & +31:12:39 & 9 & 0.83$\pm$0.11 & 0.63 &
11.16$\pm$0.16 & 1700$\pm$150 & $<$8.35 \\
Koposov 43 & FSR 848 & 05:52:14.6 & +29:55:09 & 8 & 0.38$\pm$0.10 & 0.44 &
12.21$\pm$0.09 & 2800$\pm$120 & 9.30$\pm$0.1 \\
Koposov 49 & Teutsch 10 & 05:44:22.2 & +28:49:13 & 6 & 0.42$\pm$0.05 & 0.94 &
11.46$\pm$0.22 & 2000$\pm$200 & $<$9.15 \\
Koposov 52 & Teutsch 5 & 05:53:48.9 & +26:50:26 & 5 & 1.03$\pm$0.04 & 0.58 &
12.32$\pm$0.11 & 2900$\pm$140 & 8.95$\pm$0.1 \\
Koposov 53 & & 06:08:56.2 & +26:15:49 & 3 & 0.34$\pm$0.04 & 0.61 &
12.52$\pm$0.03 & 3200 $\pm$100 & $<$8.5 \\
Koposov 62 & & 06:18:02.0 & +24:42:38 & 6 & 0.34$\pm$0.02 & 0.57 &
12.21$\pm$0.05 & 2800$\pm$60  & 9.40$\pm$0.1 \\
Koposov 63 & FSR 869 & 06:10:01.7 & +24:33:38 & 5 & 0.26$\pm$0.04 & 0.40 &
12.32$\pm$0.28 & 3000$\pm$350 & 9.15$\pm$0.1 \\
Koposov 77 &  & 05:43:52.3 & +21:42:37 & 5 & 0.57$\pm$0.01 & 0.55 &
11.23$\pm$0.02 & 1750$\pm$50  & 9.65$\pm$0.1 \\
\end{tabular}
\end{table*}

\begin{table}
\caption{Coordinates of new infrared embedded clusters} 
\label{new_IR_OC}
\centering
\begin{tabular}{llccc}
\hline\hline
{Name} & 
{Other name} &
{RA (J2000)} & 
{Dec (J2000)} & 
{D}
\\
 &
 &
{h:m:s} & 
{d:m:s} & 
{arcmin}
\\ \hline
Koposov 7 & FSR 784 & 05:40:44.1 & +35:55:25 & 6   \\
Koposov 41 & FSR 839 & 06:03:58.0 & +30:15:41 & 4   \\
Koposov 58 & FSR 849 & 05:51:11.0 & +25:46:41 & 2   \\
Koposov 82 & Teutsch 136 & 06:11:55.8 & +20:40:14 & 4    \\
\end{tabular}
\end{table}

Also, we found 4 new infrared clusters embedded in the nebulae, which are
similar to the clusters found by \citet{bica2003a} and \citet{bica2003b}: their
coordinates are presented in Table~\ref{new_IR_OC}, and
Hess-diagrams are shown in Fig.\ref{k7_k41}-\ref{k58_82} in the Appendix. 
For IR clusters, the cluster
reveals itself as a cloud on the left-hand diagram; the effect of the
differential extinction is also clearly noticeable. The right-hand diagram
displays CMD of field stars around the cluster. Because of a high value of
reddening, it was impossible to fit isochrones and find parameters for
these clusters, except Koposov 41. The Hess-diagram for this cluster exhibits
the main sequence, and we can fit it with the isochrone of the age of 4
Myr. This best-fit isochrone was used to find the distance of 2200
pc and the color excess of $E(B-V) = 1.95$.

\begin{figure*}
\centering
\includegraphics[width=17cm]{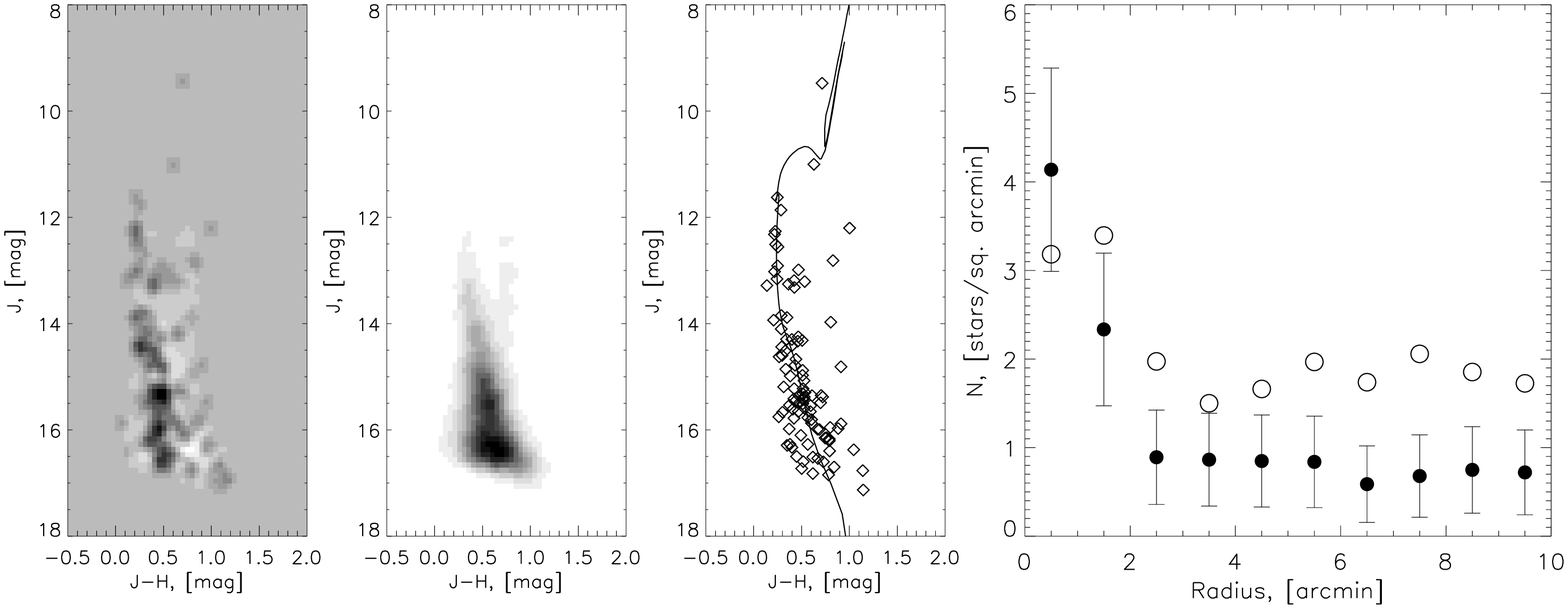}
\caption{First and second columns: Hess diagram
of the Koposov 10 cluster and Hess diagram of the
background. Third column: CMD diagram of the stars within $2'$ from the
center of the Koposov 10 with the fitted
isochrone. Fourth column: Radial density distribution for Koposov 10, the
symbols used are the same as in Fig.~\ref{radial_distrib_koposov52}.}
\label{k10}
\end{figure*}

\begin{figure*}
\centering
\includegraphics[width=17cm]{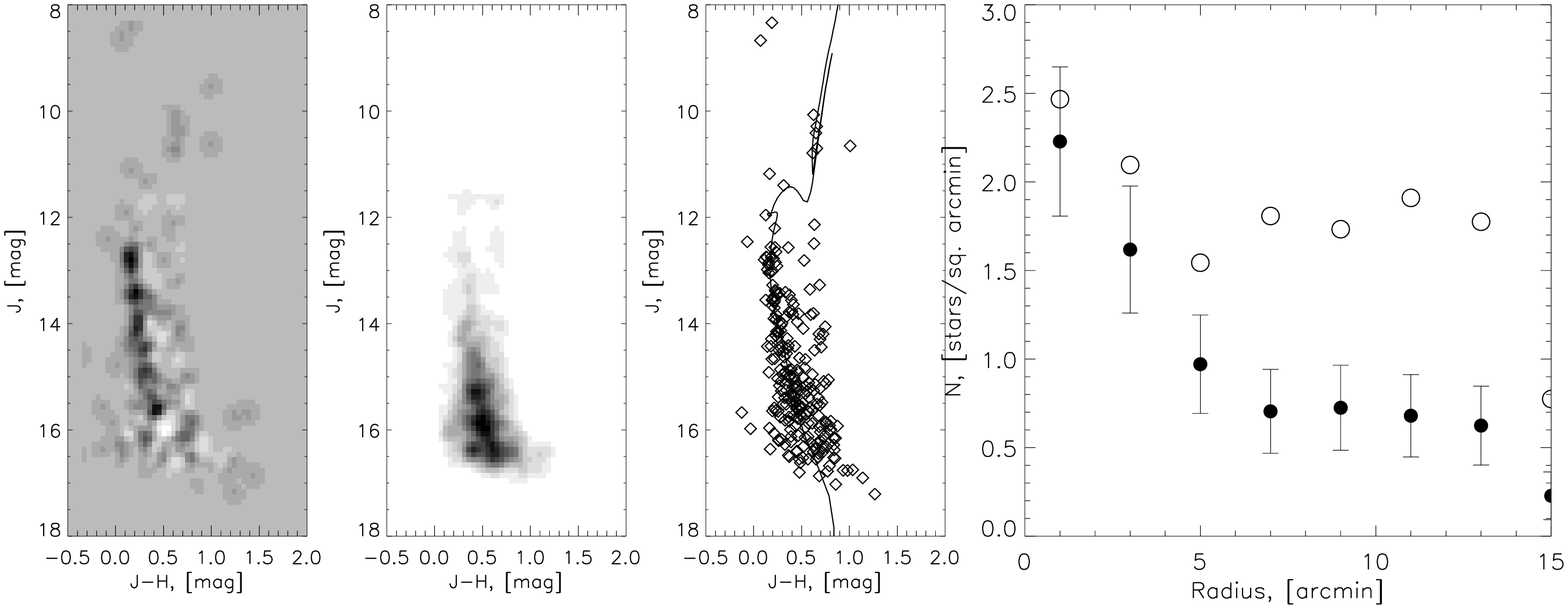}
\caption{First and second columns: Hess diagram
of the Koposov 12 cluster and Hess diagram of the
background. Third column: CMD diagram of the stars within $4'$ from the
center of the Koposov 12 with the fitted
isochrone. Fourth column: Radial density distribution for Koposov 12, the
symbols used are the same as in Fig.~\ref{radial_distrib_koposov52}.}
\label{k12}
\end{figure*}

\begin{figure*}
\centering
\includegraphics[width=17cm]{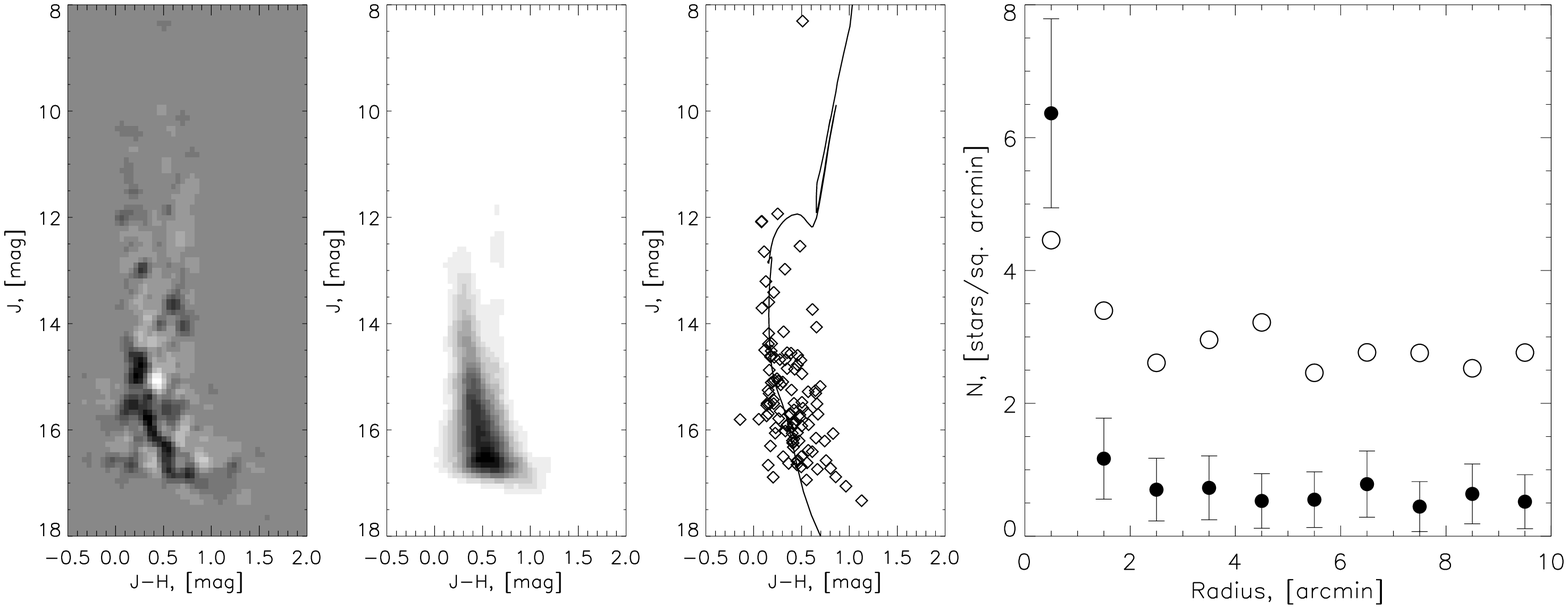}
\caption{First and second columns: Hess diagram
of the Koposov 27 cluster and Hess diagram of the
background. Third column: CMD diagram of the stars within $2\arcmin$ from
the center of the Koposov 27 with the fitted
isochrone. Fourth column: Radial density distribution for Koposov 27, the
symbols used are the same as in Fig.~\ref{radial_distrib_koposov52}.}
\label{k27}
\end{figure*}

\begin{figure*}
\centering
\includegraphics[width=17cm]{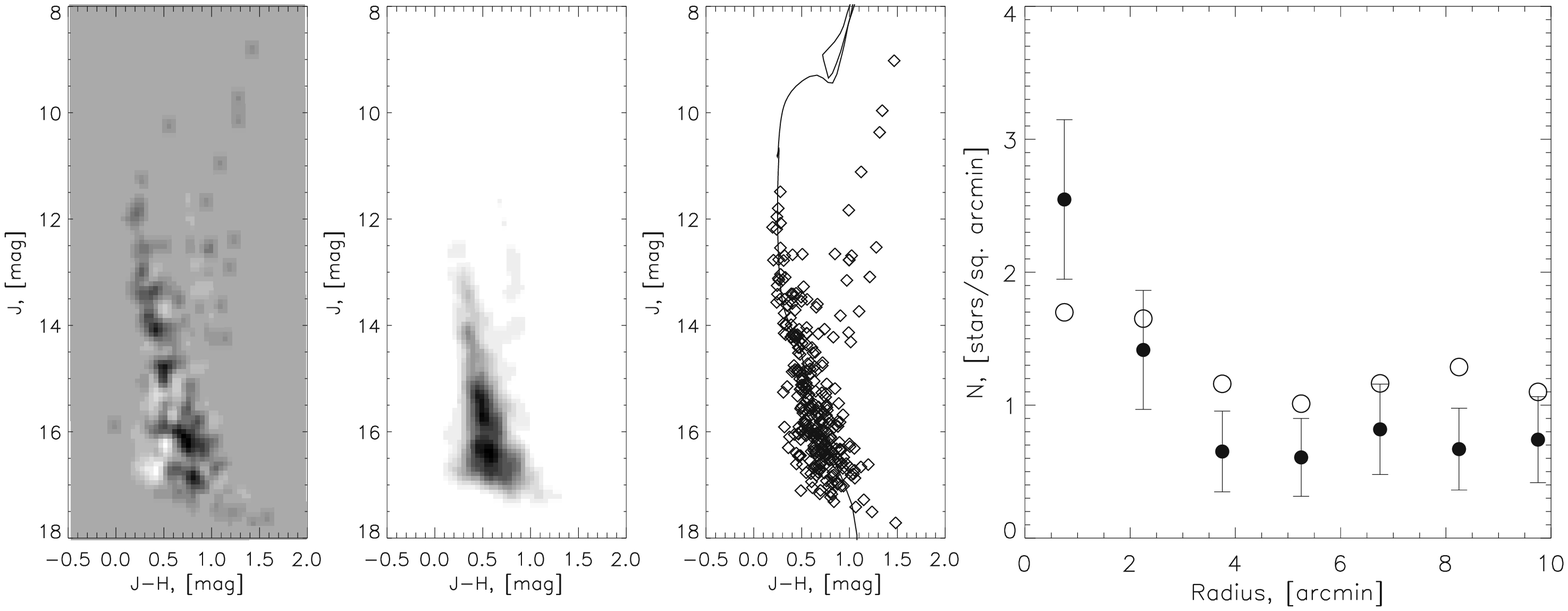}
\caption{First and second columns: Hess diagram
of the Koposov 36 cluster and Hess diagram of the
background. Third column: CMD diagram of the stars within $4\arcmin$ from
the center of the Koposov 36 with the fitted
isochrone. Fourth column: Radial density distribution for Koposov 36, the
symbols used are the same as in Fig.~\ref{radial_distrib_koposov52}.}
\label{k36}
\end{figure*}

\begin{figure*}
\centering
\includegraphics[width=17cm]{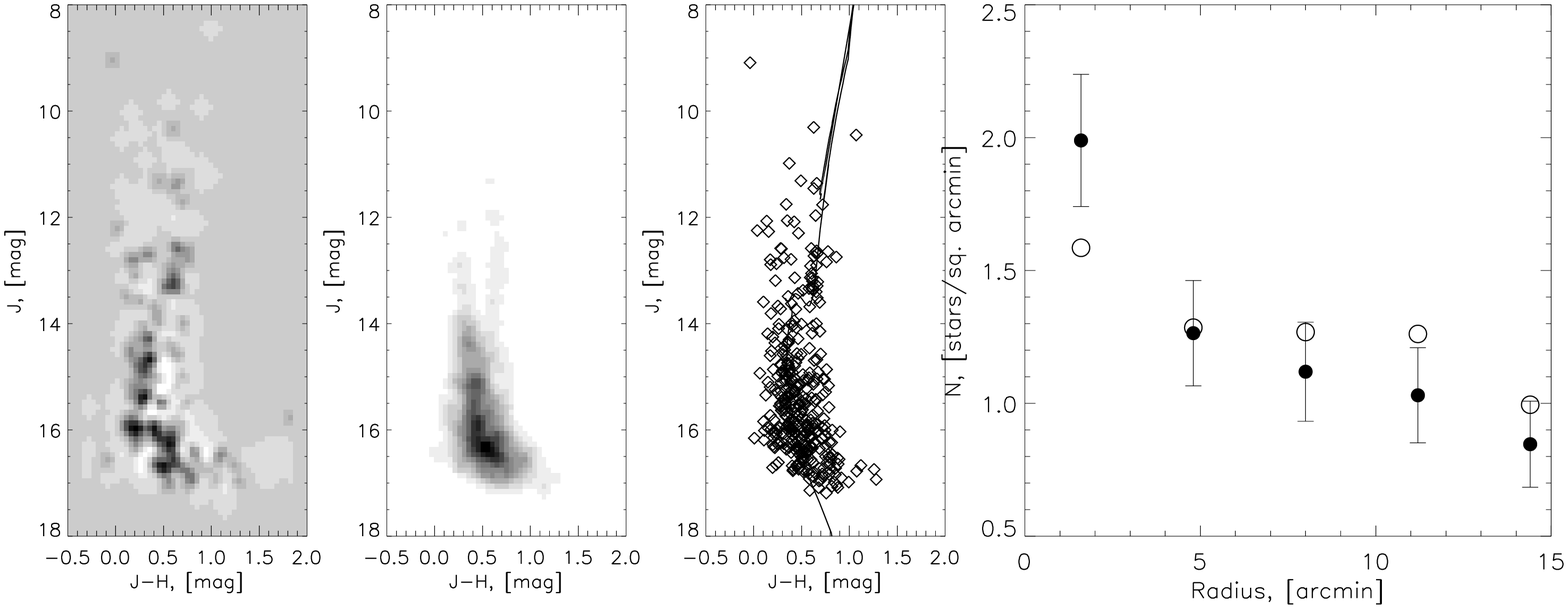}
\caption{First and second columns: Hess diagram
of the Koposov 43 cluster and Hess diagram of the
background. Third column: CMD diagram of the stars within $4\arcmin$ from
the center of the Koposov 43 with the fitted
isochrone. Fourth column: Radial density distribution for Koposov 43, the
symbols used are the same as in Fig.~\ref{radial_distrib_koposov52}.}
\label{k43}
\end{figure*}

\begin{figure*}
\centering
\includegraphics[width=17cm]{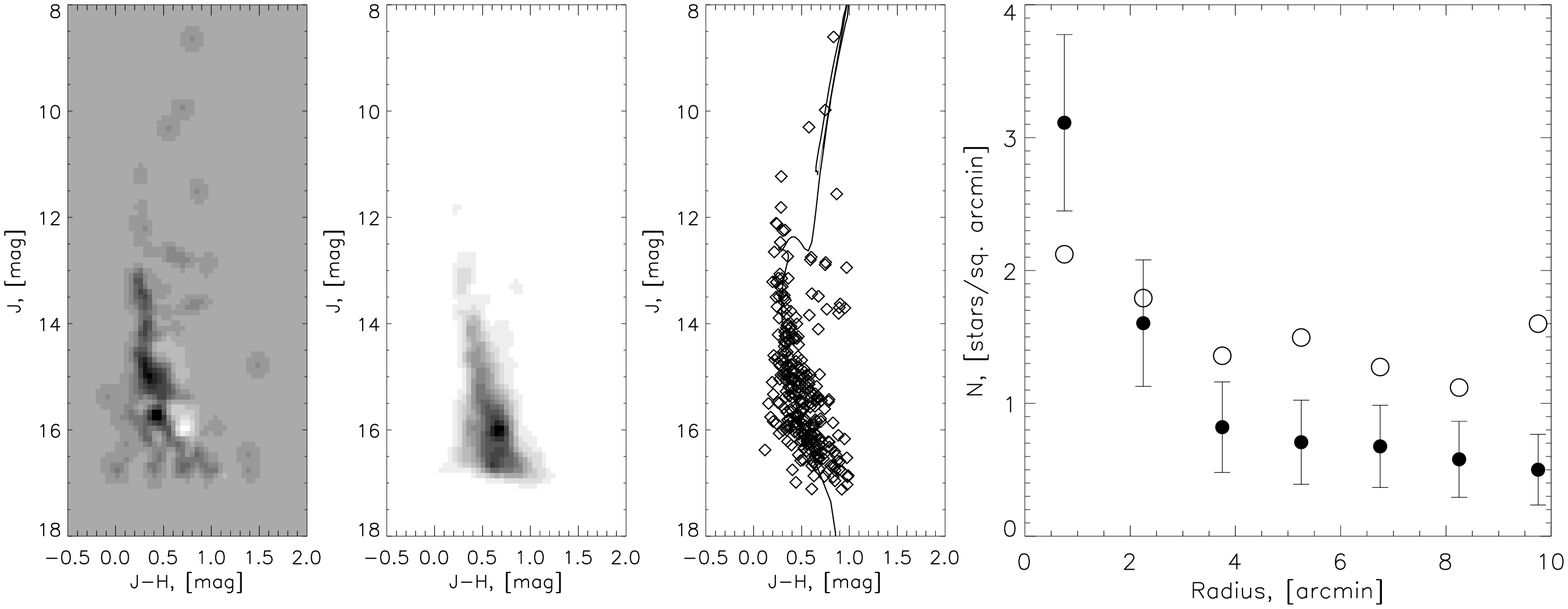}
\caption{First and second columns: Hess diagram
of the Koposov 49 cluster and Hess diagram of the
background. Third column: CMD diagram of the stars within $4\arcmin$ from
the center of the Koposov 49 with the fitted
isochrone. Fourth column: Radial density distribution for Koposov 49, the
symbols used are the same as in Fig.~\ref{radial_distrib_koposov52}.}
\label{k49}
\end{figure*}

\begin{figure*}
\centering
\includegraphics[width=17cm]{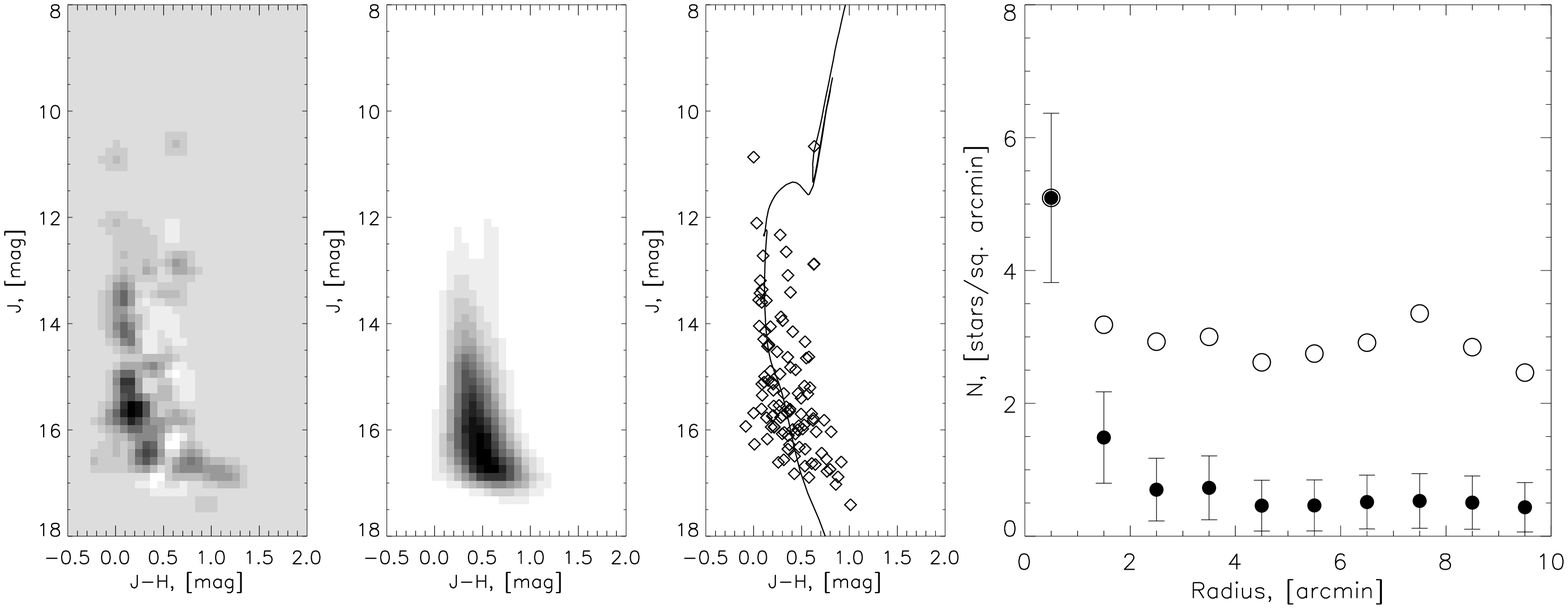}
\caption{First and second columns: Hess diagram
of the Koposov 53 cluster and Hess diagram of the
background. Third column: CMD diagram of the stars within $4\arcmin$ from
the center of the Koposov 53 with the fitted
isochrone. Fourth column: Radial density distribution for Koposov 53, the
symbols used are the same as in Fig.~\ref{radial_distrib_koposov52}.}
\label{k53}
\end{figure*}

\begin{figure*}
\centering
\includegraphics[width=17cm]{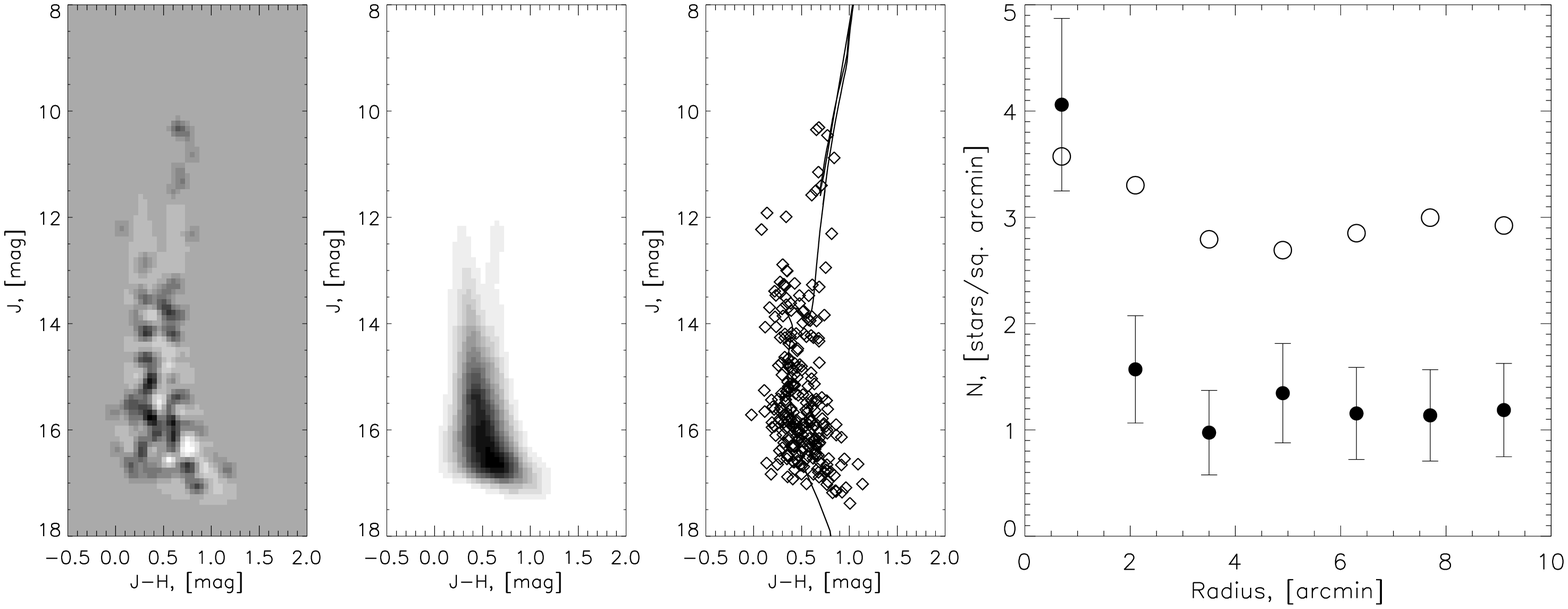}
\caption{First and second columns: Hess diagram
of the Koposov 62 cluster and Hess diagram of the
background. Third column: CMD diagram of the stars within $3\arcmin$ from
the center of the Koposov 62 with the fitted
isochrone. Fourth column: Radial density distribution for Koposov 62, the
symbols used are the same as in Fig.~\ref{radial_distrib_koposov52}.}
\label{k62}
\end{figure*}

\begin{figure*}
\centering
\includegraphics[width=17cm]{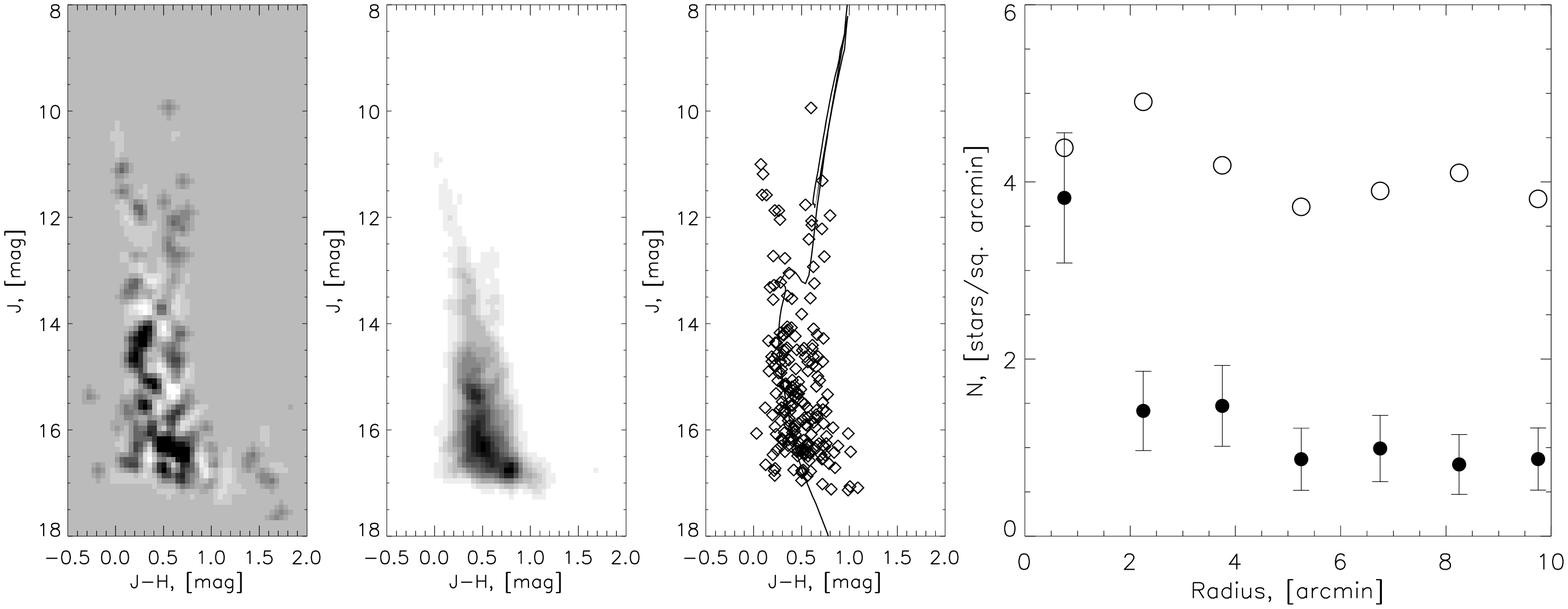}
\caption{Koposov 63 open cluster. First and second columns: Hess diagram
of the Koposov 63 cluster and Hess diagram of the
background. Third column: CMD diagram of the stars within $2\arcmin.5$ from
the center of the Koposov 63 with the fitted
isochrone. Fourth column: Radial density distribution for Koposov 63, the
symbols used are the same as on figure~\ref{radial_distrib_koposov52}.}
\label{k63}
\end{figure*}

\begin{figure*}
\centering
\includegraphics[width=17cm]{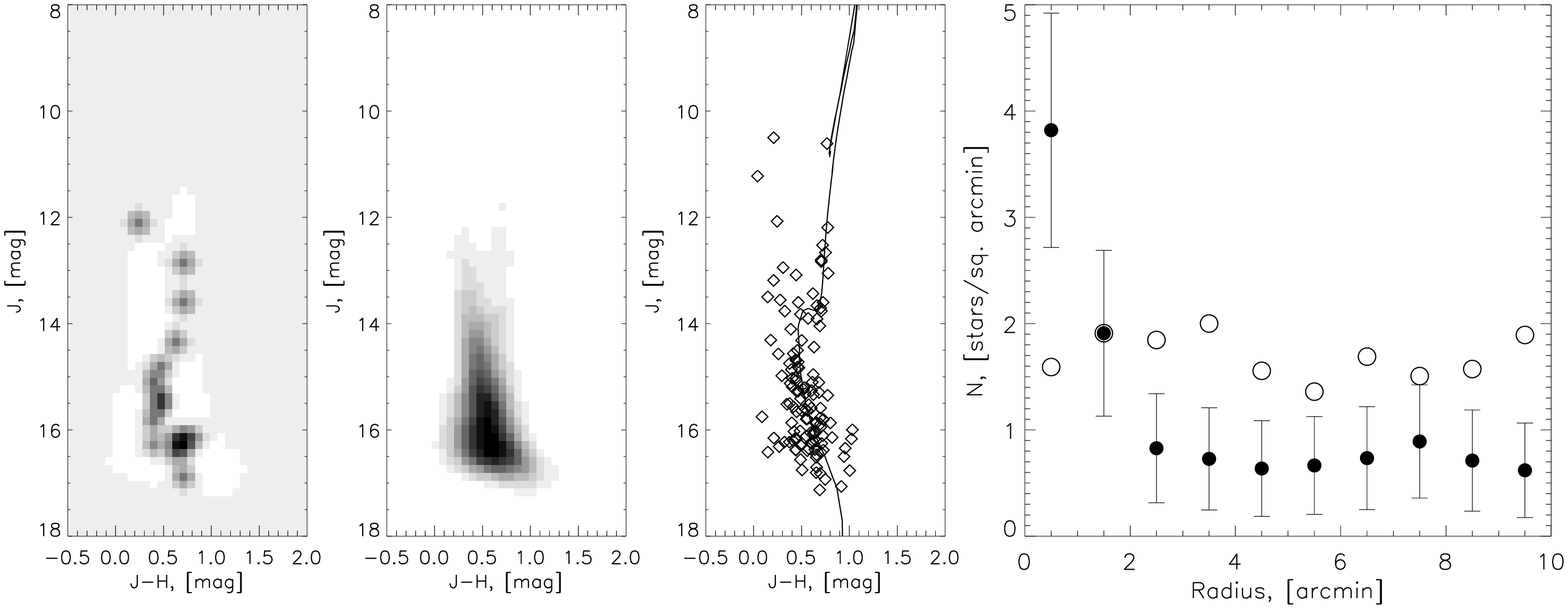}
\caption{Koposov 77 open cluster. First and second columns: Hess diagram
of the Koposov 77 cluster and Hess diagram of the
background. Third column: CMD diagram of the stars within $2\arcmin.5$ from
the center of the Koposov 77 with the fitted
isochrone. Fourth column: Radial density distribution for Koposov 77, the
symbols used are the same as on figure~\ref{radial_distrib_koposov52}.}
\label{k77}
\end{figure*}



\begin{figure*}
\begin{center}
\includegraphics[width=9cm]{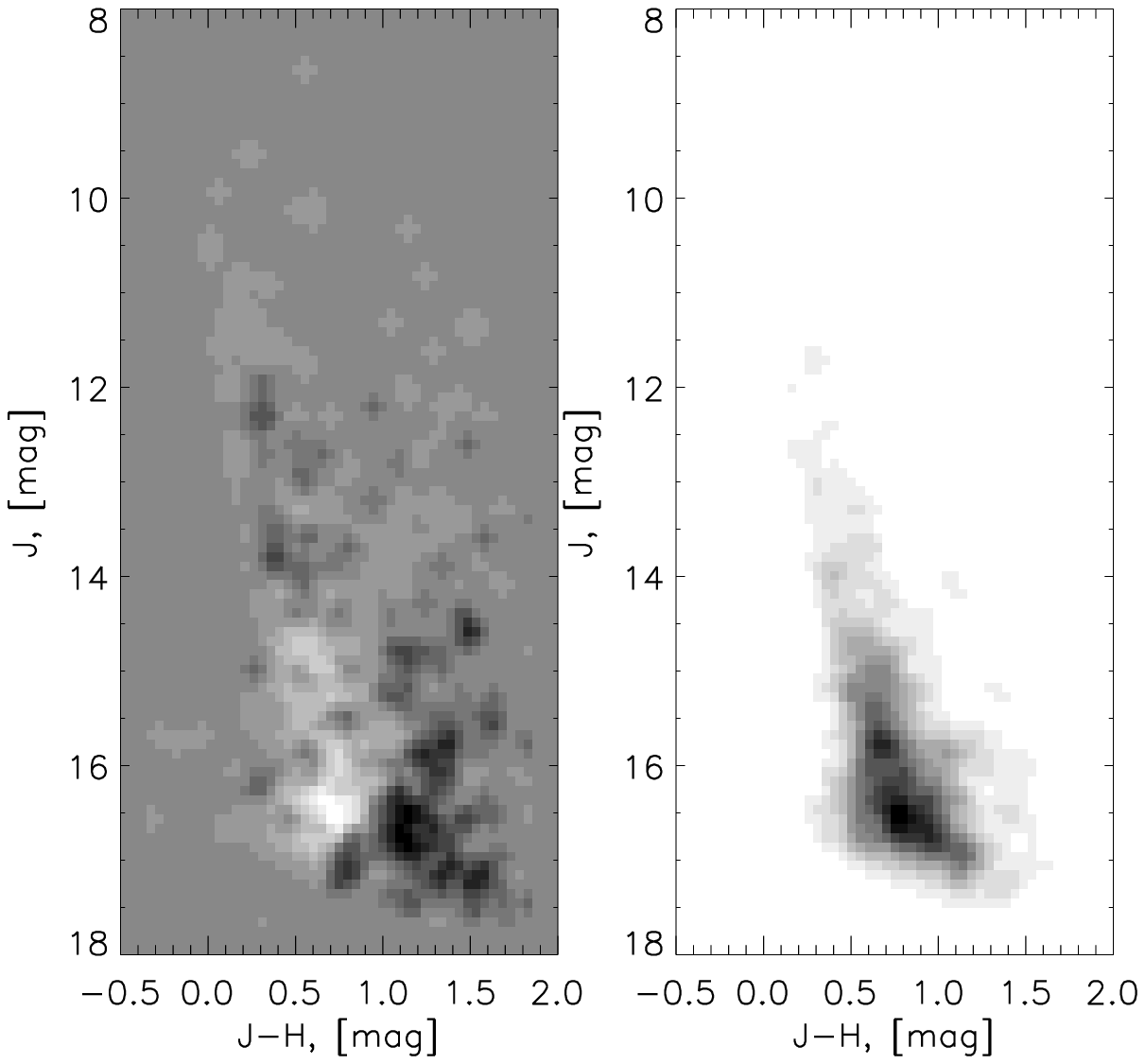}
\includegraphics[width=9cm]{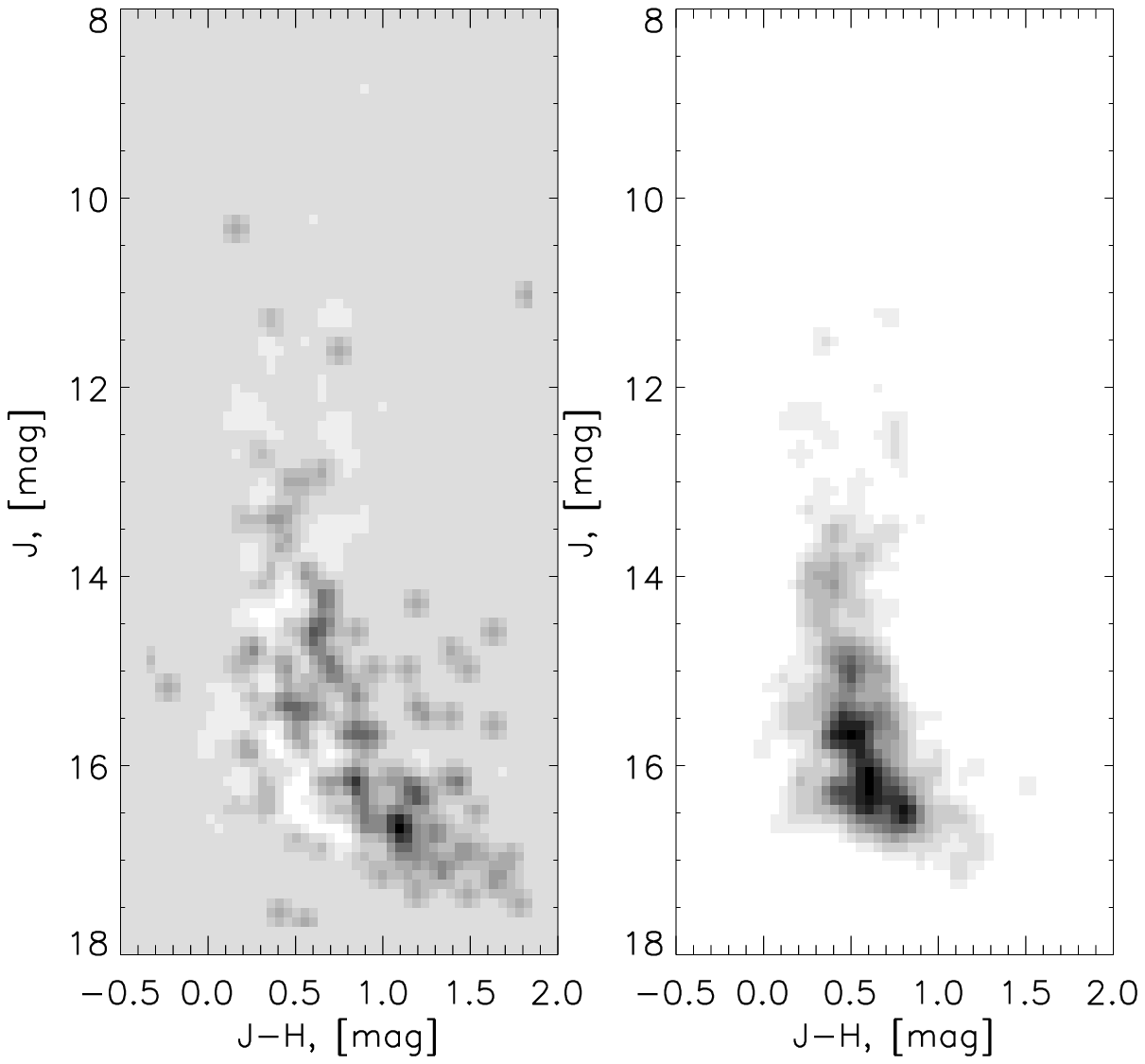}
\caption{Left panel: Hess diagram of the IR-embedded cluster Koposov 7 and Hess
diagram of the background. Right panel: Hess diagram of the IR-embedded cluster
Koposov 41 and Hess diagram of the background}
\label{k7_k41}
\end{center}
\end{figure*}



\begin{figure*}
\begin{center}
\includegraphics[width=9cm]{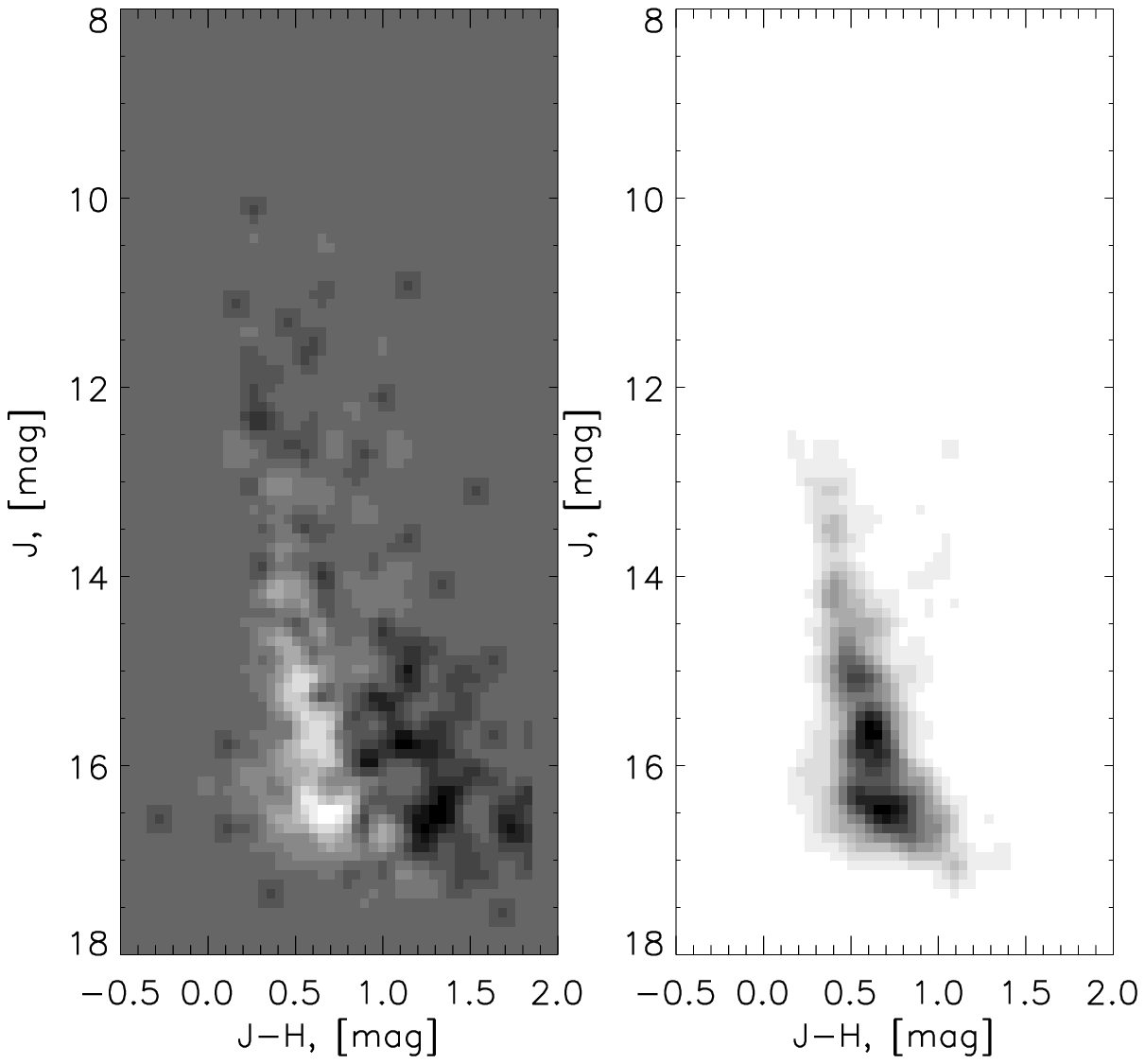}
\includegraphics[width=9cm]{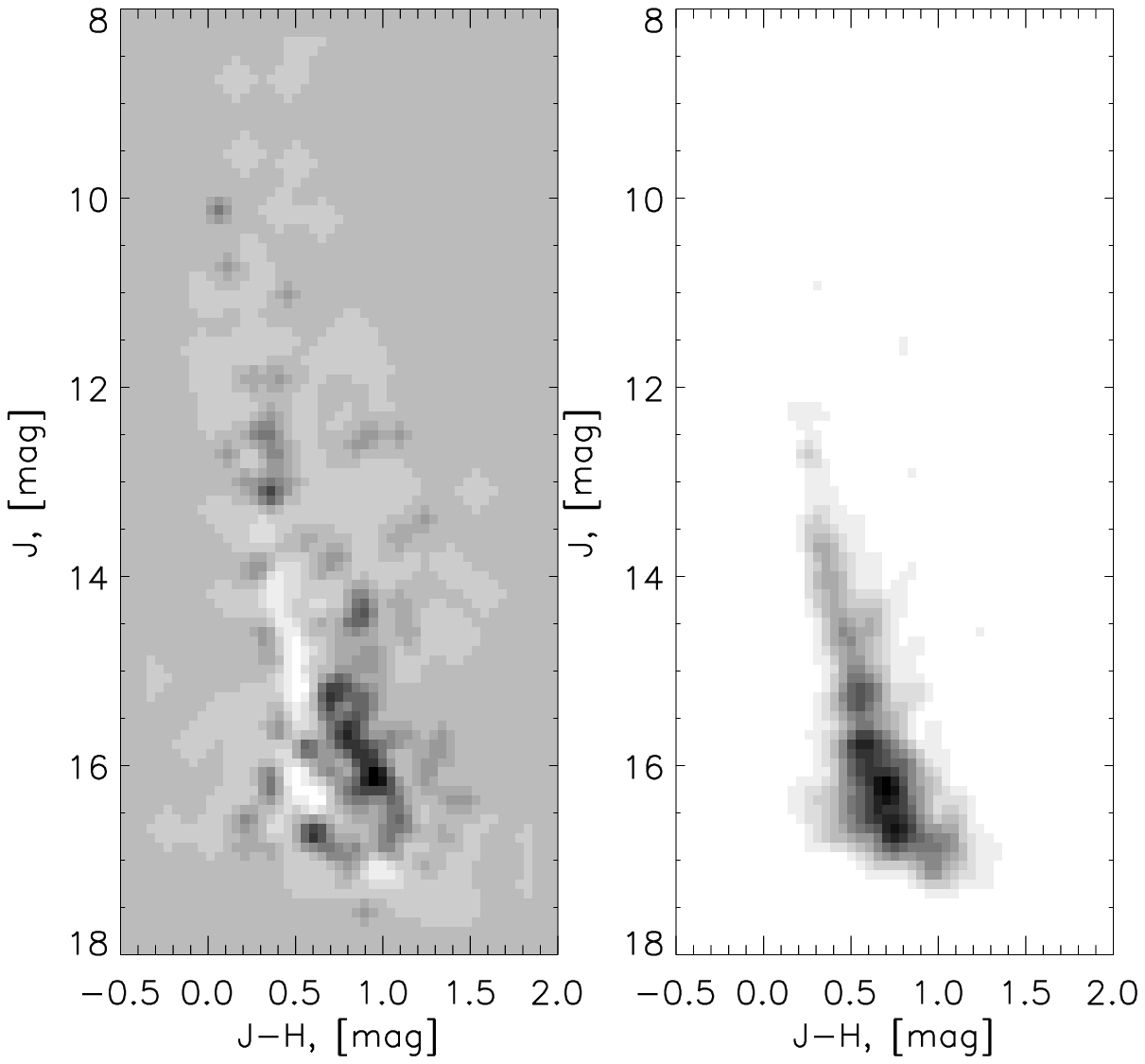}
\caption{Left panel: Hess diagram of the IR-embedded cluster Koposov 58 and Hess
diagram of the background. Right panel: Hess diagram of the IR-embedded cluster
Koposov 82 and Hess diagram of the background}
\label{k58_82}
\end{center}
\end{figure*}

Thirty two overdensities turned out to be known clusters: 23 were matched
to the
objects from the Dias catalog \citep{dias2002}, and 9, to IR clusters
from the list by \citet{bica2003a} and \citet{bica2003b}.

In Table~\ref{we_dias_comparison}, we present the data for all clusters from the
catalog compiled by \citet{dias2002}, that were detected by our technique
within the square region studied.
Although two clusters, NGC 1912 and NGC 2168, were not detected because
of their large diameters (about 25 arcmin), we added their parameters to the
table, as they exhibit slight overdensities and can be studied by our methods. In some cases, we obtained more precise coordinates of the center of
clusters, so we give new coordinates for all clusters. 

\begin{table*}
\begin{center}
\caption{Comparison of physical parameters of the clusters: Dias catalog
and Paunzen-Mermilliod database (DPM) vs. present study (KGZ)}
\label{we_dias_comparison}
\begin{tabular}{lcccccccc}
\hline
\hline
{Name} & {RA(J2000)} &
{Dec(J2000)} & {$D_{DPM}$} & {$D_{KGZ}$} &
{$E(B-V)_{DPM}$} & {$E(B-V)_{KGZ}$} & {$Age_{DPM}$} &
{$Age_{KGZ}$} \\
{}& {h:m:s} & {d:m:s} & {pc} &
{pc} & {mag} & {mag} & {log(yr)} & 
{log(yr)}
\\
\hline
Basel 4       & 05:48:54.9 & +30:11:08 & 3000 & 2750 & 0.45  & 0.57 & 8.30  & 8.25 \\
Berkeley 17   & 05:20:29.6 & +30:34:33 & 2700 & 2400 & 0.58  & 0.30 & 10.00 & 10.00 \\
Berkeley 19   & 05:24:02.8 & +29:34:16 & 4831 & 3000 & 0.40  & 0.61 & 9.49  & 9.25 \\ 
Berkeley 21   & 05:51:47.4 & +21:48:31 & 5000 & 5150 & 0.76  & 0.51 & 9.34  & 9.35 \\
Berkeley 69   & 05:42:22.6 & +22:50:01 & 2860 & 2900 & 0.65  & 0.45 & 8.95  & 9.00 \\
Berkeley 71   & 05:40:56.7 & +32:16:33 & 3900 & 2450 & 0.85 & 0.91 & 8.80 & 8.80 \\
Berkeley 72   & 05:50:17.6 & +22:14:59 & {}   & 3500 & {} & 0.43 & {} & 8.65 \\
Czernik 21    & 05:26:41.0 & +36:00:49 &  {}  & 2300 & {} & 0.72 & {} & 9.55 \\
Czernik 23    & 05:50:03.6 & +28:53:41 & {} & 2500 & {} & 0.38 & {} & 8.45 \\  
Czernik 24    & 05:55:24.6 & +20:53:11 &  {}   & 4600 & {}  & 0.26 & {} & 9.40 \\
DC 8          & 06:09:21.3 & +31:13:54 &  {}  & 2100 &  {}  & 0.72 & {} & 9.00 \\ 
IC 2157       & 06:04:41.9 & +24:06:01 & 2040 & 2400 & 0.548 & 0.58 & 7.800
& $<$7.6 \\ 
King 8        & 05:49:19.0 & +33:37:38 & 6403 & 3100 & 0.580 & 0.44 & 8.618 &
9.05 \\
Kronberger 1  & 05:28:22.0 & +34:46:24 & 1900 &  800 & 0.52  & 0.43 & 7.5   &
8.10 \\
NGC 1893      & 05:22:53.7 & +33:26:17 & 6000 & {} & 0.45  & {} & 6.48  & {}  \\
NGC 1907      & 05:28:10.7 & +35:19:44 & 1800 & 1300 & 0.52  & 0.51 & 8.5   & 8.60 \\
NGC 1931      & 05:31:25.9 & +34:12:50 & 3086 & 1000 & 0.738 & 1.97 & 7.002 & $<$7.0 \\
NGC 1960      & 05:36:19.6 & +34:07:27 & 1330 & 1050 & 0.22  & 0.19 & 7.4   & $<$7.5 \\
NGC 2099      & 05:52:18.4 & +32:33:03 & 1383 & 1300 & 0.302 & 0.27 & 8.540 & 8.60 \\
NGC 2129      & 06:01:10.5 & +23:19:34 & 2200 & 1950 & 0.80  & 0.82 & 7.00  & 7.10 \\  
NGC 2158      & 06:07:27.8 & +24:05:53 & 5071 & 3300 & 0.360 & 0.34 & 9.023 & 9.30 \\
Pismis 27     & 06:10:53.8 & +20:36:26 & {} & 1000 & {} & 0.68 & {} &
$<$7.7 \\
Stock 8       & 05:28:08.8 & +34:25:53 & 1821 &  900 & 0.445 & 1.21 & 7.056 & $<$7.5 \\
NGC 1912      & 05:28:41.6 & +35:48:34 & 1400 & 1000 & 0.25  & 0.38 & 8.5   &
8.30 \\
NGC 2168      & 06:09:00.0 & +24:21:00 &  912 &  900 & 0.20  & 0.19 & 8.25  &
7.95 \\
\end{tabular}
\end{center}
\end{table*}

In the columns, we give parameters of the clusters both listed in the Dias
catalog \citep{dias2002} or in WEBDA database and measured by applying
our methodology to the
2MASS catalog. For
clusters Be 72, Cz 21, Cz 23, Cz 24, DC 8, Pis 27, Dias does not
publish any parameters except their coordinates and diameters. However, in
the WEBDA database there are data on Be 71 taken from
the paper by \citet{lata2004}: $E(B-V)=0.85$, $d=3900 pc$, $log(t)=8.80$.
We observe that the color excess and the age well agree with our
parameters, whereas the distance exceeds our value by 1500 pc. The
difference in distance evaluations (when the age estimations are pretty close)
can be attributed to the fact that the authors fitted $(V,B-V)$ CMD by Zero
Age Main Sequence (ZAMS) 
given by \citet{schmidt-kaler1982} to estimate the distance modulus, and the 
theoretical isochrones given by \citet{girardi2002} to find the cluster's 
age, while we used 2MASS theoretical isochrones by \citet{girardi2002}
to evaluate all cluster's parameters.
 
Parameters we measured for 7 clusters (Be 19, Be 71, Kronberger 1, King
8, NGC
1931, Stock 8 and NGC 2158) significantly differ from the data in the catalog
by \citet{dias2002} and in WEBDA, but we suppose our results are more precise and 
homogeneous. Parameters of Be 19 were obtained by 
\citet{christian1980} as following: employing
the UBV photographic photometry data, the author compared the CMD of Be 19
with both the color-magnitude diagram of NGC 752 and the theoretical
isochrones by \citet{ciardullo1979}. 
Parameters of King 8 cluster, which appeared in the
catalog by \citet{dias2002} were calculated by \citet{loktin2001}
using photographic photometry by \citet{christian1981}. However, \citet{christian1981}
published the distance to the clusters equal to 3.5 kpc, which essentially
differs from the value of 6403 pc by \citet{loktin2001}.
Parameters of Stock 8 were also obtained by \citet{loktin2001}
using the UBV photometry of 32 stars only measured by different
authors; these stars do not show any sequence on CMD. As for cluster
Kronberger 1, the isochrone corresponding to the parameters available in the
\citet{dias2002} catalog fitted the main sequence of field stars,
whereas the cluster members are noticeably shifted to the right in the
$(J,J-H)$ CMD because of extinction. The effect of extinction is clearly
seen on the corresponding Hess-diagrams. 
Parameters of NGC 1931 cluster were obtained by \citet{loktin2001} using
published photometry data by \citet{bhatt1994}, but the
authors themselves
give the distance of 2170 pc. This cluster is embedded in a nebula, so
its Hess-diagram shows a slightly scattered area occupied by the cluster.
We fitted its CMD with a very young isochrone, and the distance
appeared to be three times smaller than that by \citet{dias2002}.
The estimations of the distance to NGC 2158 cluster
differ from each other very noticeably (by 2 kpc). \citet{dias2002}
take data from \citet{loktin2001}, who used compilative data 
and automatic method to find the cluster's parameters.   
At the same time, our data are in good accordance with parameters by \citet{carraro2002}
 who obtained 3600 pc for the distance, $E(B-V)=0.55$, and $log(t)= 9.3$ by
fitting isochrone from \citet{girardi2000} and then comparing the best-fit
with the simulated synthetic CMD.
The discussion above demonstrates that the authors often publish controversial
estimations of the cluster parameters even if they employ the same observational
data. Although our estimations are sometimes based on less deep photometric
data, they have an essential advantage in homogeneity of both observational data
and used isochrones and fitting methods.

Members of NGC 1893 cluster detected on the Hess-diagram do not lie on the
same isochrone, so we cannot find a satisfactory fitting. This fact can be
reasonably attributed to the existence of pre-main sequence stars in this
cluster described by \citet{vallenari1999}. The distance to the cluster
found by \citet{vallenari1999} equals 4300 pc (cf. with 6000 pc in the
Dias catalog).

In the remaining 11 cases, our parameters closely agree (to within the
error
levels) with the
corresponding data available from the catalog by \citet{dias2002}. These
error levels are as follow: 200-500 pc for the distances, $0.10^m$ for the
color excess $E(B-V)$, and 0.05 for $log(t)$. 

There are five additional clusters that have parameters in
Dias' catalog and that were not detected in the field under study (see
Sect. 3). However, the distances and other parameters for four of them were
obtained from very poor (V,B-V) diagrams, which do not show any sequence or
clump
at the CMD, and we did not consider these clusters. 

At the end, 41 of 88 density peaks showed no evidence of being real clusters. 
In these cases, we detect either a nebula, or a bright star, or occasional 
groups of stars, whose density for some reasons exceeds the density of field 
stars.

In Fig.~\ref{clusters_distribution}, we also show the distribution of
all clusters under investigation across the galactic plane: both the
newly-opened clusters and the known ones, whose parameters were (re)determined
in the
present study. The crosses indicate the clusters with $log(age)$ less
then 8.00; all young clusters are situated from 1 to 3.5 kpc from the Sun.

We do not quantify the detection efficiency of our
algorithm. In the square under our study in the vicinity of the Galactic
anticenter, we detect as much as $95\%$ of clusters listed with the
diameters
less than 15 arcminutes in the \citet{dias2002} catalog. But ideally the
efficiency can be determined only with simulations, since none of existing
catalags is complete and free of selection effects. The detailed analysis of
the detection efficiency and selection effects of the algorithm is left for our
future work.

\section{Conclusions}

We have demonstrated the new method of searching for star clusters in
the
data from large surveys and applied it to the 2MASS data.
In the small field of 16 by 16 degrees in the poor region of Galactic
anticenter, we have found and verified 15 new open clusters. Ten
of them coincide with cluster candidates from papers by
\citet{froebrich2007} and \citet{kronberger2006}. However, we not only
discovered these objects independently, but also investigated their nature. We
developed an automated method, which involves three different techniques:
Hess-diagram, color-magnitude diagram in $(J,J-H)$ and $(K_s,J-K_s)$, and radial
density distribution.  For 12 of the new clusters,
we obtained main physical parameters: ages, distances, and color excesses. We
also used our methods to evaluate the same physical parameters for all known
clusters detected in the selected area. We found that out of 25 such clusters,
only in 11 cases we can accept the values of the cluster parameters
listed in
the catalog by \citet{dias2002}. We have found, or improved, the distances,
ages, and color excesses for 13 previously-known clusters. 
As one can see from  Fig.~\ref{clusters_distribution}, 
the number of clusters in the square studied,
for which their main physical parameters are reliably evaluated, has increased
from 11 in the catalog by \citet{dias2002} to 35 in our study.

Therefore, to get a catalog
of homogeneously-measured parameters of open clusters, it is necessary not only
to search for new clusters and thoroughly investigate them, but also to
recalculate the parameters of all known clusters using uniform raw datasets and
a uniform and automated processing methodology.

\begin{acknowledgements}

The work was supported by the Russian Foundation for Basic Research (grant no.
08-02-00381) and the President Grant NSH-5290.2006.2.
S. Koposov is supported by the DFG through SFB 439 and by a EARA-EST Marie
Curie Visiting fellowship.

This research has made use of the SAI Catalog Access Services,
Sternberg Astronomical Institute, Moscow, Russia.

This publication makes use of
data products from the Two Micron All Sky Survey,
which is a joint project of the University of Massachusetts and the Infrared
Processing and Analysis Center/California Institute of Technology, funded by the
National Aeronautics and Space Administration and the National Science
Foundation. 

We thank for Hans-Walter Rix, Vasily Belokurov and Wilton Dias for the comments
on earlier versions of this paper.
\end{acknowledgements}

\end{document}